\documentclass[prd,superscriptaddress,amsfonts,amssymb,amsmath,showpacs,twocolumn]{revtex4-1}
\usepackage{bm}
\usepackage{amsfonts}
\usepackage{latexsym}
\usepackage[latin1]{inputenc}
\usepackage{graphicx}
\usepackage{amsmath}
\usepackage{palatino}
\usepackage{mathpazo}
\usepackage{textcomp}
\usepackage{booktabs}
\usepackage{dcolumn}
\usepackage{hyperref}
\hypersetup{colorlinks,citecolor=blue}
\usepackage{amsmath}
\usepackage{xcolor}

\def\jnl@style{\it}
\def\aaref@jnl#1{{\jnl@style#1}}

\def\aaref@jnl#1{{\jnl@style#1}}

\def\aj{\aaref@jnl{AJ}}                   
\def\apj{\aaref@jnl{ApJ}}                 
\def\apjl{\aaref@jnl{ApJ}}                
\def\apjs{\aaref@jnl{ApJS}}               
\def\apss{\aaref@jnl{Ap\&SS}}             
\def\aap{\aaref@jnl{A\&A}}                
\def\aapr{\aaref@jnl{A\&A~Rev.}}          
\def\aaps{\aaref@jnl{A\&AS}}              
\def\mnras{\aaref@jnl{Mon.~Not.~Roy.~Astron.~Soc.}}             
\def\prd{\aaref@jnl{Phys.~Rev.~D}}        
\def\prc{\aaref@jnl{Phys.~Rev.~C}}  
\def\prl{\aaref@jnl{Phys.~Rev.~Lett.}}    
\def\qjras{\aaref@jnl{QJRAS}}             
\def\skytel{\aaref@jnl{S\&T}}             
\def\ssr{\aaref@jnl{Space~Sci.~Rev.}}     
\def\zap{\aaref@jnl{ZAp}}                 
\def\nat{\aaref@jnl{Nature}}              
\def\aplett{\aaref@jnl{Astrophys.~Lett.}} 
\def\apspr{\aaref@jnl{Astrophys.~Space~Phys.~Res.}} 
\def\physrep{\aaref@jnl{Phys.~Rep.}}      
\def\physscr{\aaref@jnl{Phys.~Scr}}       
\def\commat{\aaref@jnl{Comm.~Math.~Phys.}}              
\def\science{\aaref@jnl{Science}}               
\def\cqg{\aaref@jnl{Classical Quant.~Grav.}}            
\def\jpcs{\aaref@jnl{JPCS}}                                     
\def\ijmpd{\aaref@jnl{Int.~J.~Mod.~Phys.~D}}                    
\def\grg{\aaref@jnl{Gen.~Relat.~Gravit.}}               
\def\rpp{\aaref@jnl{Rep.~Prog.~Phys.}}          
\def\npa{\aaref@jnl{Nucl.~Phys.~A}}        
\def\lrr{\aaref@jnl{Living Rev.~Rel.}}                   
\def\jcap{\aaref@jnl{J.~Cosmology Astropart.~Phys.}}    
\def\rmp{\aaref@jnl{Rev.~Mod.~Phys.}}   

\allowdisplaybreaks[1]

\begin{document}
\color{red}

\title{Nonlinear magnetically charged black holes in 4D Einstein-Gauss-Bonnet gravity}
\author{Kimet Jusufi}
\email{kimet.jusufi@unite.edu.mk}
\affiliation{Physics Department, State University of Tetovo,  Ilinden Street nn, 1200, Tetovo, North Macedonia}
\affiliation{Institute of Physics, 
  Faculty of Natural Sciences and Mathematics,
  Ss. Cyril and Methodius University,
  Arhimedova 3, 1000 Skopje, North Macedonia}

\begin{abstract}
In this letter we present an exact spherically symmetric and magnetically charged black hole solution with exponential model of nonlinear electrodynamics [S. Kruglov, Annals Phys. 378, 59-70 (2017)] in the context of  4D Einstein-Gauss-Bonnet (EGB) gravity.  
We show that our $-$ve branch, in the limit of GB coupling coefficient $\alpha \rightarrow 0$ and the nonlinear parameter $\beta \to 0$, reduces to the magnetically charged black hole of Einstein-Maxwell gravity in GR. In addition we study the embedding diagram of the black hole geometry and the thermodynamic properties such as the Hawking temperature and the heat capacity of our black hole solution.
 
\end{abstract}

\maketitle

\section{Introduction}\label{sec1}
It is well known that the EGB theory is topological in $4D$ as the $GB$ Lagrangian is a total derivative and, therefore, it does not contribute to the gravitational dynamics in $4D$. In a recent work Glavan \& Lin \cite{Glavan:2019inb} proposed an idea based on the rescaling the Gauss-Bonnet coupling constant $\alpha$ as $\alpha/(D -4)$, and taking the limit $D \to 4$ at the level of the field equation to obtain in this way a non-trivial contribution in $4D$. This novel 4D EGB gravity has interesting properties such as bypasses the conclusions of Lovelock's theorem and avoids Ostrogradsky instability.
Furthermore the static and spherically symmetric vacuum black holes found in \cite{Glavan:2019inb} have interesting properties, for example the gravitational force is repulsive at short
distance and thus an infalling particle never reaches
$r=0$ point. In other words, the theory is
free from singularity problem. This is in contrast to
Einstein's general relativity, where an infalling particle
will eventually hit the singularity and effective theory
breaks down, this is also the case in $HD$ black holes \cite{Boulware:1985wk}.

After the work of Glavan and Lin \cite{Glavan:2019inb} the 4D EGB theory received compelling attention. The charged AdS black hole was obtained in Ref. \cite{Fernandes:2020rpa}, black holes in the four-dimensional Einstein-Lovelock gravity, \cite{Konoplya:2020qqh}, clouds of string in the novel $4D$ EGB black holes \cite{Singh:2020nwo}, a Vaidya metric Ref.~\cite{Ghosh:2020vpc}, generating black holes solution was also addressed in Ref. \cite{Ghosh:2020syx}, Hayward and Bardeen black holes in 4D EGB theory \cite{Kumar:2020xvu,Kumar:2020uyz}, rotating black holes using Newman-Janis algorithm \cite{Kumar:2020owy,Wei:2020ght}, rotating black hole as particle accelerator \cite{NaveenaKumara:2020rmi}, thermodyanmical properties of AdS black hole were studied in Ref. \cite{HosseiniMansoori:2020yfj}, QNMS, stability and shadows  \cite{Konoplya:2020bxa}, gravitational lensing by black holes \cite{Islam:2020xmy}, strong gravitational lensing in homogeneous plasma \cite{Jin:2020emq}, stability of the Einstein Static Universe in $4 D$ EGB \cite{Li:2020tlo}, QNMs and Strong Cosmic Censorship \cite{Mishra:2020gce}, wormholes in 4D EGB \cite{Jusufi:2020yus}, thin shell wormholes \cite{thinshell}, relativistic stars  \cite{Doneva:2020ped}, 4D EGB as heat engine \cite{1}, the innermost stable circular orbit and shadow \cite{2}, greybody factor and power spectra of the Hawking radiation in the novel $4D$ EGB de-Sitter gravity \cite{3}, superradiance and stability of the novel 4D charged EGB black hole  \cite{4},
weak cosmic censorship conjecture for the novel $4D$ charged EGB black hole with test scalar field and particle  \cite{5}, extended thermodynamics and microstructures in AdS space  \cite{6}, spinning test particle in 4D EGB \cite{7}, perturbative and nonperturbative QNMs of 4D EGB \cite{8}, regularized Lovelock gravity  \cite{9},
thin accretion disk around 4D EGB \cite{10} and other interesting studies \cite{11,12,13,14,15,16}.

In the same time, objections on the 4D EGB theory were reported in the work of Gurses et al.  \cite{Gurses:2020ofy} and Ref. \cite{Ai:2020peo}. Importantly, in a very recent works, it was argued that a well-defined $D \to 4$ limit of EGB gravity can be obtained and a regularized field equations has been reported in Refs. \cite{Fernandes:2020nbq,Hennigar:2020lsl}.  More specifically in Ref.  \cite{Hennigar:2020lsl} authors employed the Mann-Ross method \cite{Mann} and found that the limit $D\to 4$ is a special case of the scalar-tensor theory of the Horndeski type obtained by a dimensional reduction method. In addition it is pointed out that the
spherically symmetric spacetimes in $4D$ the black hole solution should remain valid in these regularised theories, however by going
beyond the sphericaly symmetric cases the solutions are not valid.  

In this paper we aim to find an exact black hole solution  in the context of nonlinear electrodynamics supported with magnetic charge with a lagrangian density proposed in Ref.  \cite{Kruglov:2017fck}. This model was subsequently used in Ref. \cite{ali}, while in Ref. \cite{habib} a different lagrangian has been used to obtain regular magnetic black holes. The paper is organized as follows:  In  Sec. 2, we solve the field equations in the novel 4D EGB gravity to obtain magnetically charged black hole solution.   In Sec. 3,  we  explore embedding diagram. In Sec. 4, we study the Hawking temperature and the heat capacity. Finally we comment on our results in Sec. 5.

\section{Magnetically charged black holes in 4D EGB gravity}
Let us begin by writing the action in the EGB gravity in $D$-dimensions to derive the equations of motion. The gravitational action is given by
\begin{equation}\label{action}
	\mathcal{I}_{A}=\frac{1}{16 \pi}\int d^{D}x\sqrt{-g}\left[ R +\frac{\alpha}{D-4} \mathcal{L}_{\text{GB}} \right]+\mathcal{I}_{NED}
\end{equation}
in which $g$ is the determinant of the metric $g_{\mu\nu}$ while $\alpha$ is the GB coupling coefficient and has
dimensions of $[length]^2$.  The
Lagrangian density of exponential electrodynamics on the other hand reads \cite{Kruglov:2017fck}
\begin{equation}
\mathcal{L}_{NED}=-\mathcal{F}\exp \left(-\beta \mathcal{F} \right),
\end{equation}
where
\begin{equation}
\mathcal{F}=\frac{1}{4}F_{\mu \nu} F^{\mu \nu}
\end{equation}
is the Maxwell invariant with a pure magnetic field given by the 2-form
\begin{equation}
\textbf{F}=q \sin\theta d\theta \wedge d \phi.
\end{equation}
In particular the parameter
$\beta$ possesses the dimension of the $[length]^4$ and the upper bound was reported  $\beta \leq 1 \times 10^{-23}$ T$^2$ from PVLAS experiment. The term $\mathcal{L}_{\text{GB}}$ is the Lagrangian defined and is given by
\begin{equation}
	\mathcal{L}_{\text{GB}}=R^{\mu\nu\rho\sigma} R_{\mu\nu\rho\sigma}- 4 R^{\mu\nu}R_{\mu\nu}+ R^2\label{GB}.
\end{equation}
\begin{figure*}
\includegraphics[width=8.0 cm]{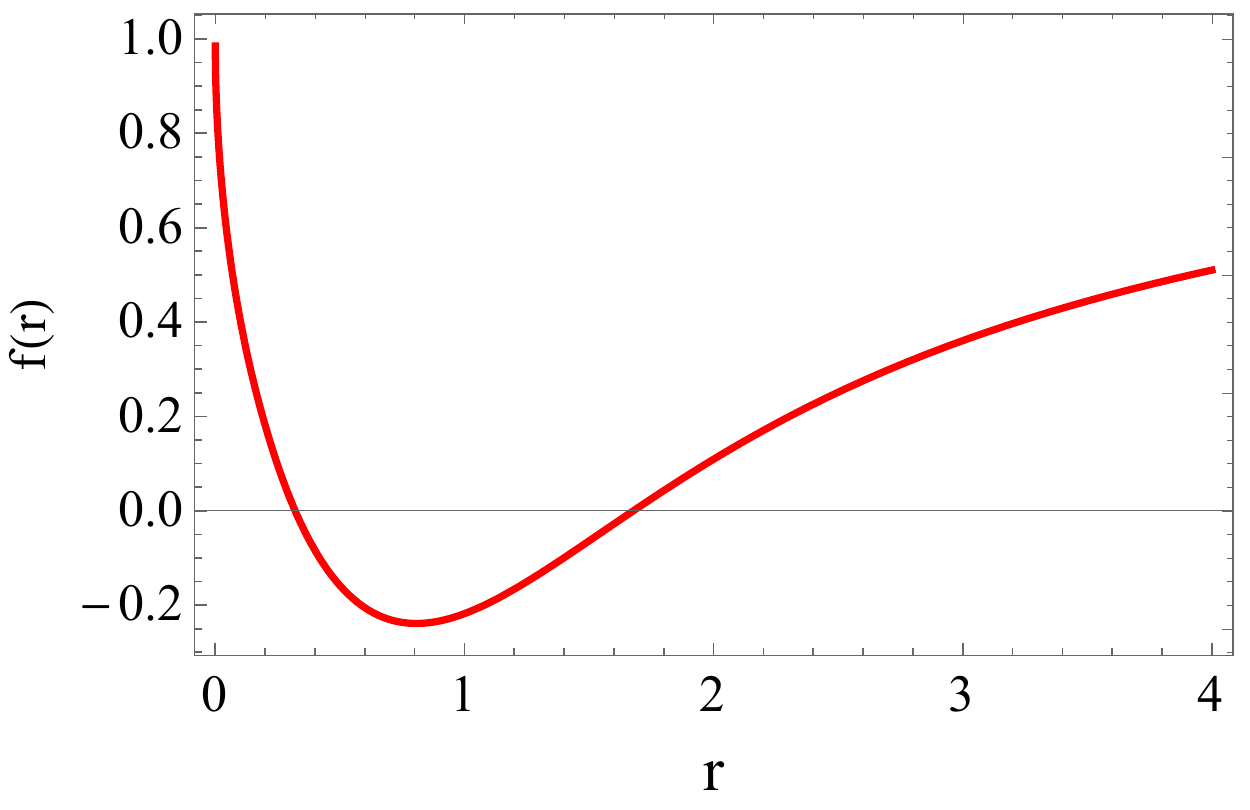}
\includegraphics[width=8.0 cm]{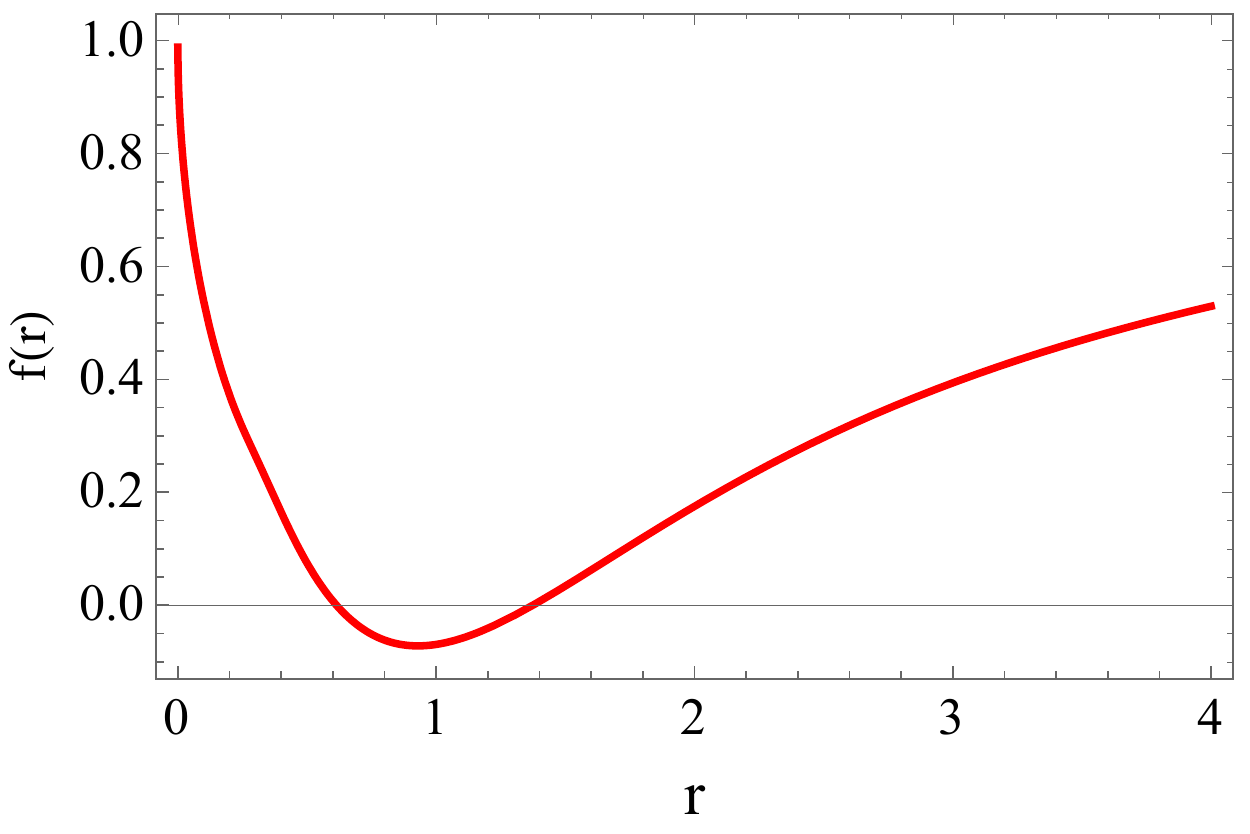}
\includegraphics[width=8.0 cm]{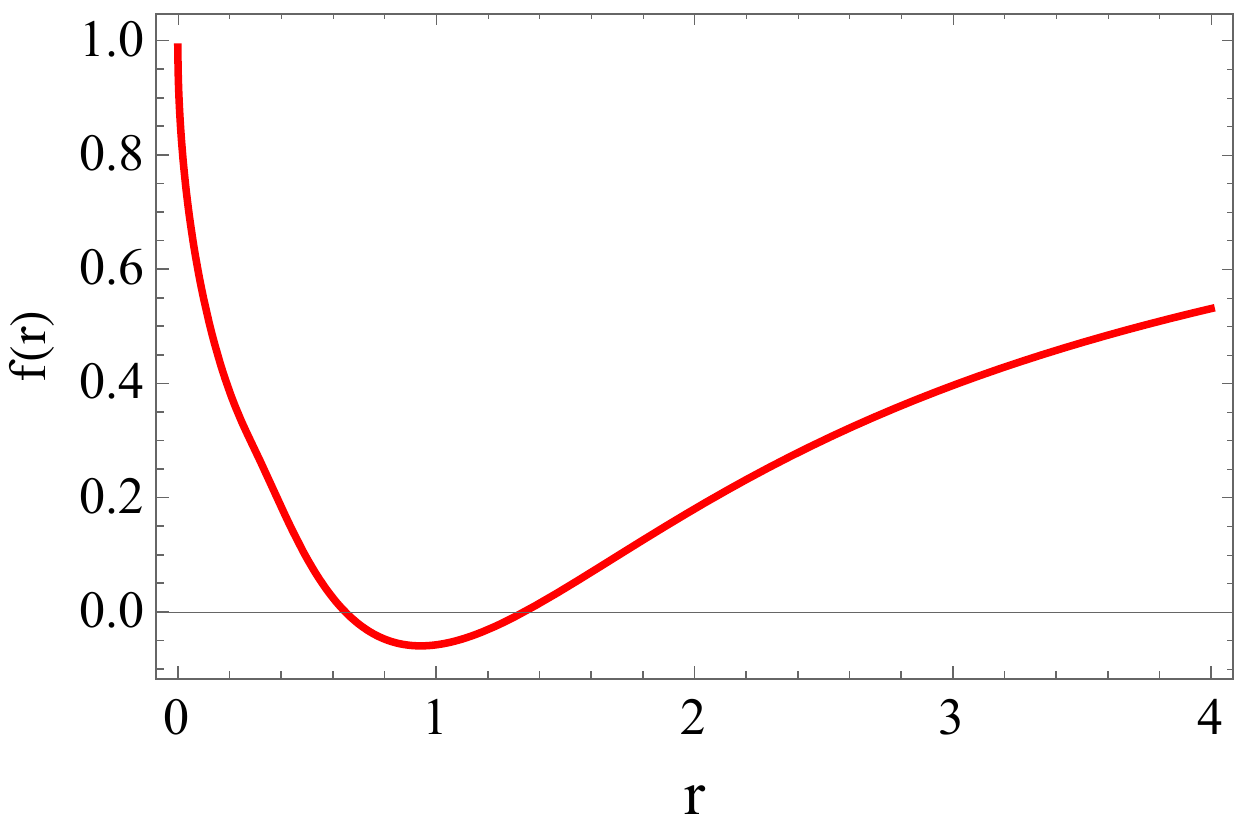}
\includegraphics[width=8.0 cm]{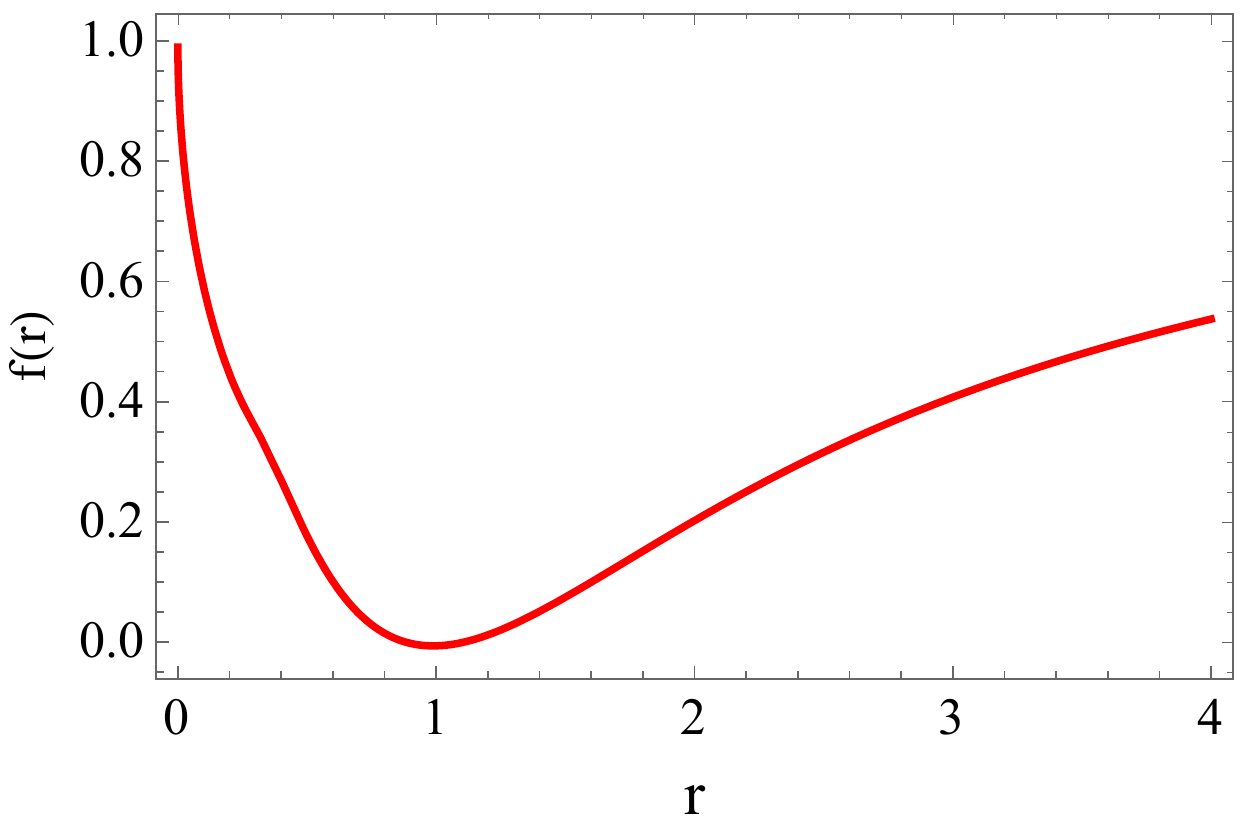}
\caption{Upper left panel: Plot for $f(r)$ as a function of $r$ for chosen $\alpha=0.5$, $\beta=0.1$, $M=1$ and $q=0.2$. Upper right panel: Plot for $f(r)$ for chosen $\alpha=0.5$, $\beta=0.1$, $M=1$ and $q=0.6$. Button left panel: Plot for $f(r)$ for chosen $\alpha=0.5$, $\beta=0.1$, $M=1$ and $q=0.62$. Button right panel: Plot for $f(r)$ for chosen $\alpha=0.5$, $\beta=0.1$, $M=1$ and $q=0.7$. Depending on the parameter values the spacetime can can have two horizons known as the Cauchy and event horizons, then an extremal black hole with degenerate horizons and finally no horizons at all.  }\label{f1}
\end{figure*}
The variation of (\ref{action}) with respect to metric $g_{\mu \nu}$ gives the field equations \cite{Ghosh:2020vpc}
\begin{equation}\label{GBeq}
	G_{\mu\nu}+\frac{\alpha}{D-4} H_{\mu\nu}= 8 \pi T^{NED}_{\mu\nu},
\end{equation}
where the energy momentum tensor in our case redas \cite{Kruglov:2017fck}
\begin{equation}
T^{\mu \nu}_{NED}=\frac{1}{4 \pi}\exp\left(-\beta \mathcal{F} \right) \left[  (1-\beta \mathcal{F})F^{\mu \lambda}F^{\nu}_{\lambda}-g^{\mu \nu} \mathcal{F} \right],
\end{equation}
along with the following expression
\begin{eqnarray}
	G_{\mu\nu}&=&R_{\mu\nu}-\frac{1}{2}R g_{\mu\nu},\nonumber\\\notag
	H_{\mu\nu}&=&2\Bigr( R R_{\mu\nu}-2R_{\mu\sigma} {R}{^\sigma}_{\nu} -2 R_{\mu\sigma\nu\rho}{R}^{\sigma\rho} - R_{\mu\sigma\rho\delta}{R}^{\sigma\rho\delta}{_\nu}\Bigl)\\
	&-& \frac{1}{2}\mathcal{L}_{\text{GB}}g_{\mu\nu}.\label{FieldEq}
\end{eqnarray}
In the these equations $R$ is the Ricci scalar, $R_{\mu\nu}$ the Ricci tensor, $R_{\mu\nu}$ is the so-called Lancoz tensor and finally $R_{\mu\sigma\nu\rho}$ the Riemann tensor.  As we already pointed out the GB term is total derivative and does not contribute to the field equations in 4D. But if we re-scaled the coupling constant $ \alpha/(D-4)$, and  considering maximally symmetric spacetimes with curvature scale ${\cal K}$ \cite{Ghosh:2020vpc}, we obtain
\begin{equation}\label{gbc}
\frac{g_{\mu\sigma}}{\sqrt{-g}} \frac{\delta \mathcal{L}_{\text{GB}}}{\delta g_{\nu\sigma}} = \frac{\alpha (D-2) (D-3)}{2(D-1)} {\cal K}^2 \delta_{\mu}^{\nu},
\end{equation}
hence one can see that the variation of the GB action does not vanish in $D=4$ due to the  re-scaled coupling constant \cite{Glavan:2019inb}. The general static and  spherically symmetric metric in $D$-dimensions reads
\begin{equation}
ds^2=-f(r)dt^2+\frac{dr^2}{f(r)}+r^2d\Omega_{D-2}^2.\label{metric}
\end{equation} 
with the unite sphere in $D$ dimensions 
\begin{equation}
d\Omega^2_{D-2} = d\theta^2_1 + \sum^{D-2}_{i=2}\prod^{i-1}_{j=1}
\sin^{2}\theta_j\;d\theta^2_i \;.\notag
\end{equation}

Using the energy-momentum for the energy
density it follows \cite{Kruglov:2017fck}
\begin{equation}
\rho=\frac{q^2}{2 r^4}\exp\left(-\frac{\beta q^2}{2 r^4} \right),
\end{equation}
where  for pure magnetic field in the spherically symmetric spacetime is given as
\begin{equation}
\mathcal{F}=\frac{q^2}{2 r^4}.
\end{equation}

\begin{figure*}[!htb]
		\includegraphics[width=8.4cm]{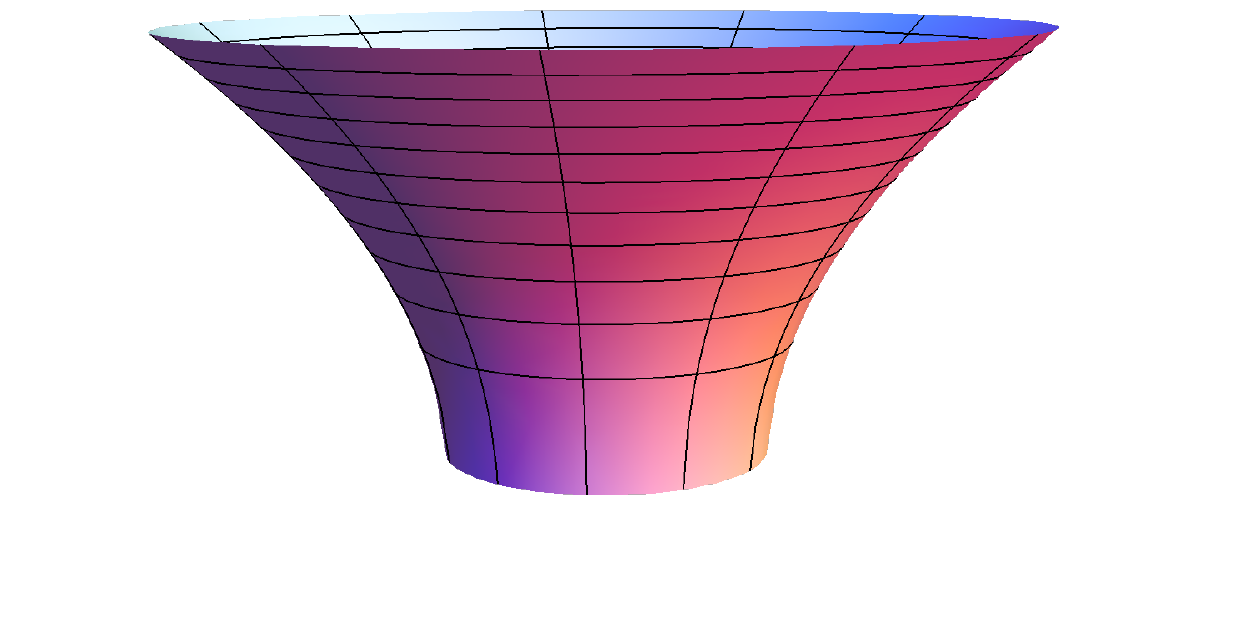}
		\includegraphics[width=8.4cm]{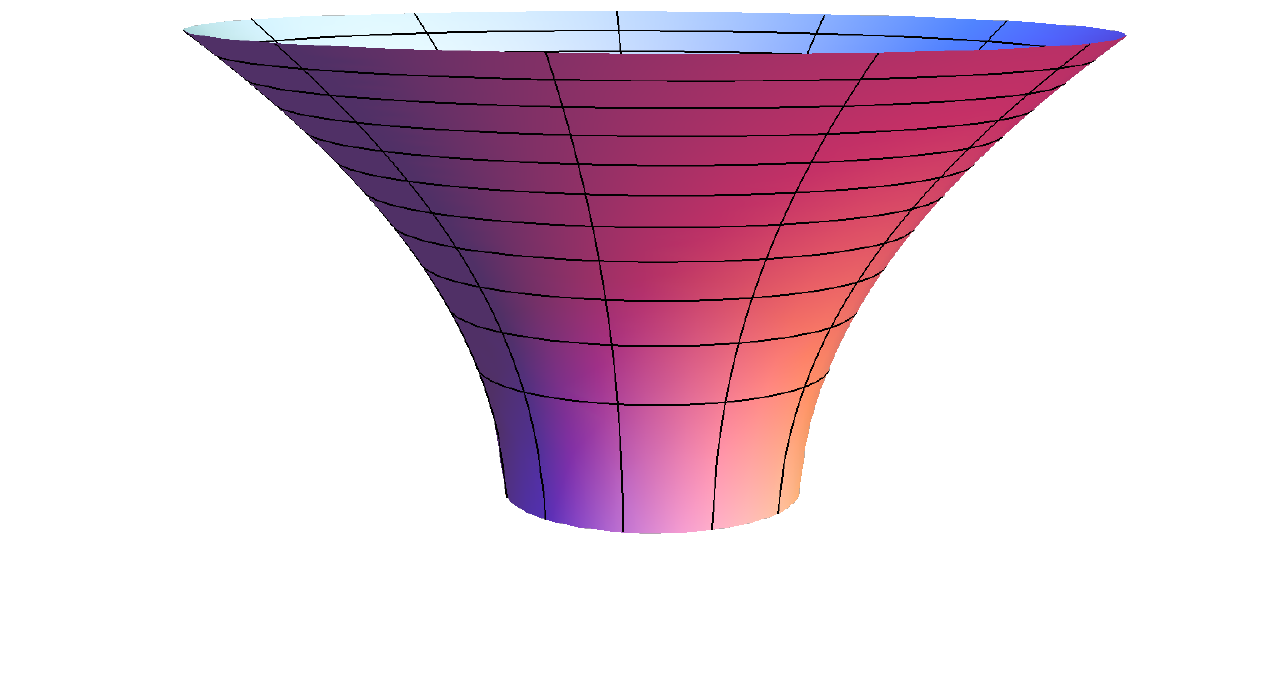}
		\caption{The BH spacetime embedded
			in a three-dimensional Euclidean space. Left panel: We  choose $M=1$, $\alpha=0.2$, $\beta=0.1$ and $q=0.2$. Right panel: We choose $M=1$, $\alpha=0.2$, $\beta=0.1$ and $q=0.6$  }
	\end{figure*}
	
The $(t-t)$ component of the Einstein field equations yields
\begin{eqnarray}\notag
&-& q^2 \exp\left(-\frac{\beta q^2}{2 r^4} \right)-r(-2\alpha f(r)+r^2+2\alpha)f'(r)\\
&-&(f(r)-1)(r^2+\alpha f(r)-\alpha)=0
\end{eqnarray}

Solving this equation we find the two branches 
\begin{equation}
f(r)=1+\frac{r^2}{2 \alpha}\left(1 \pm \sqrt{1+\frac{8 M \alpha}{r^3}-\frac{2^{13/4} \exp\left(-\frac{\beta q^2}{2 r^4} \right) \Xi q^2 \alpha }{r^4}}\right)
\end{equation}
where
\begin{equation}
\Xi=-\frac {2^{7/8} \text{WhittakerM}(\frac{1}{8},\frac{5}{8},\frac{q^2 \beta}{2r^4})) \exp\left(\frac{\beta q^2}{4 r^4} \right)+\frac{5\, 2^{3/4}}{4}(\frac{q^2 \beta}{r^4})^{1/8}}{5 (\frac{q^2 \beta}{r^4})^{1/8}}
\end{equation}

\begin{figure*}
\includegraphics[width=8.0 cm]{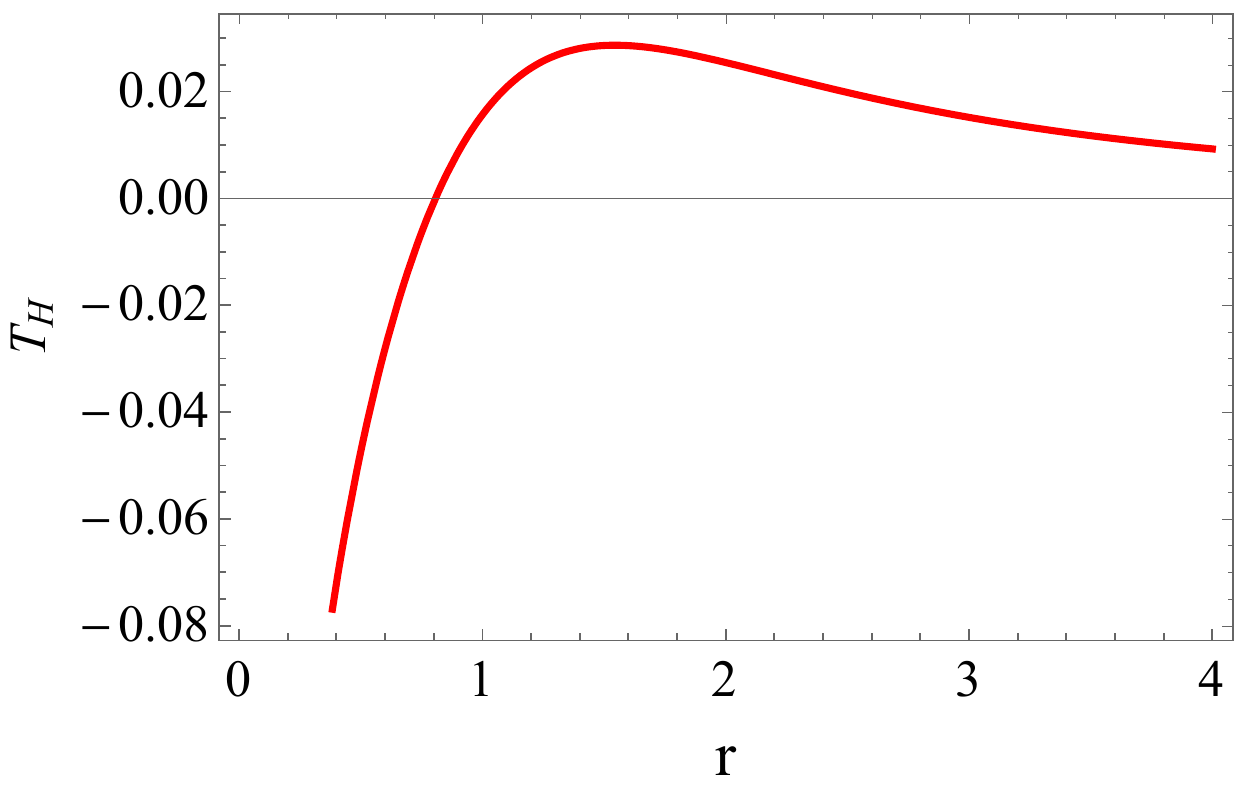}
\includegraphics[width=8.0 cm]{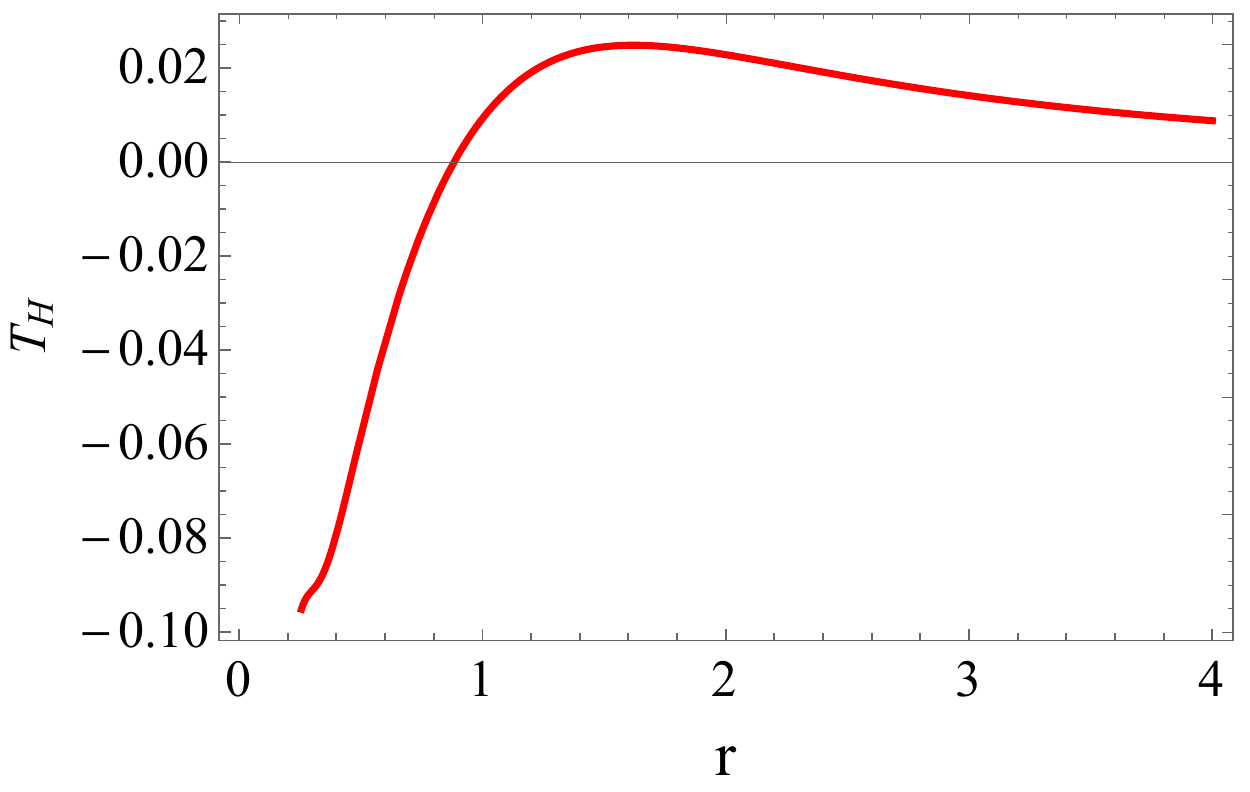}
\caption{Left panel: Plot for $T_H$ as a function of $r=r_+$ for chosen $\alpha=0.1$, $\beta=0.1$, $M=1$ and $q=0.2$. right panel: Plot for $T_H$ as a function of $r=r_+$ for chosen $\alpha=0.5$, $\beta=0.1$, $M=1$ and $q=0.5$.   }\label{f2}
\end{figure*}
 \begin{figure*}
\includegraphics[width=8.0 cm]{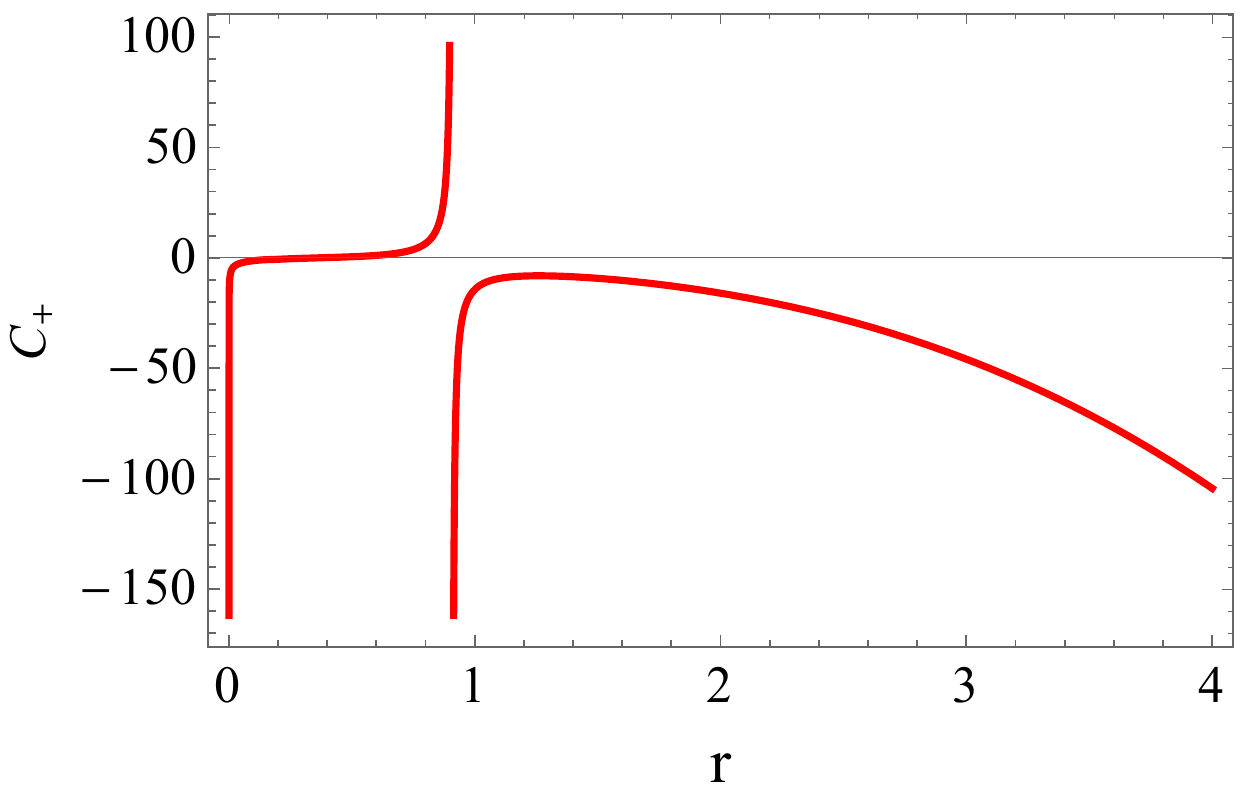}
\includegraphics[width=8.0 cm]{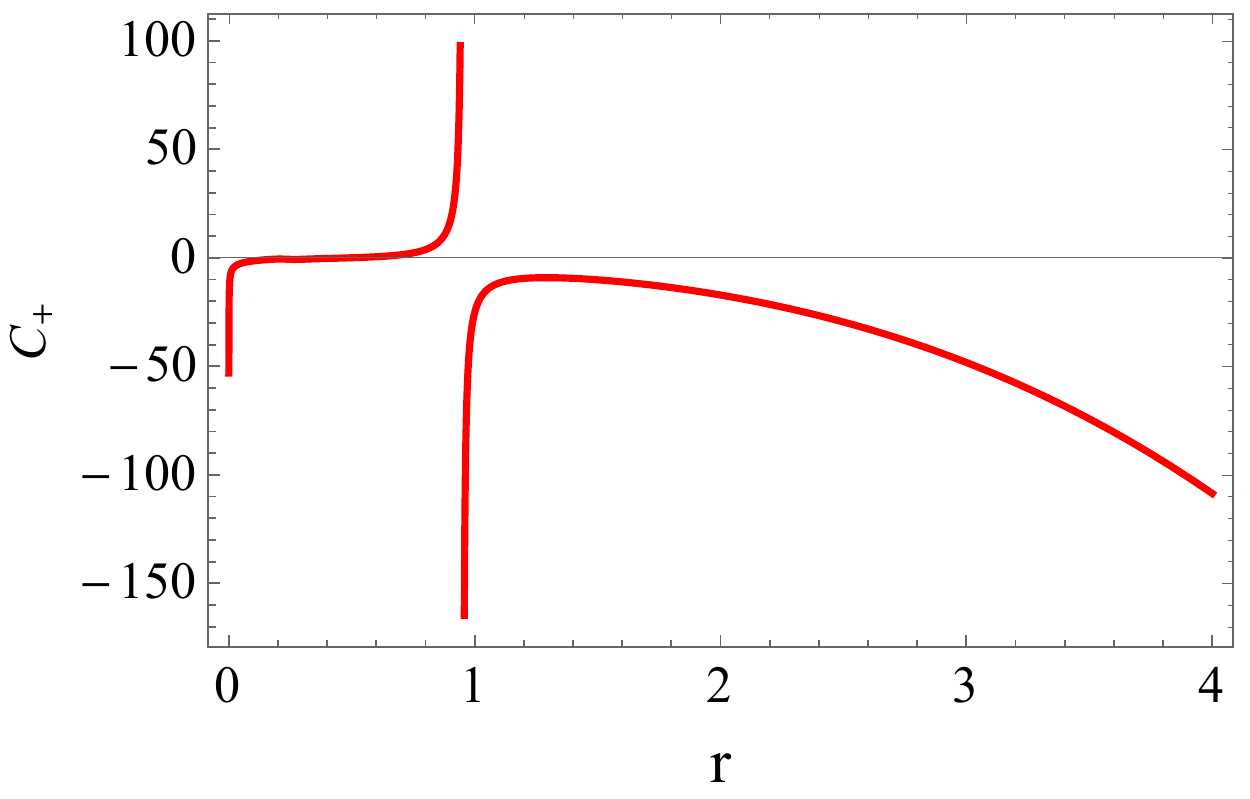}
\includegraphics[width=8.0 cm]{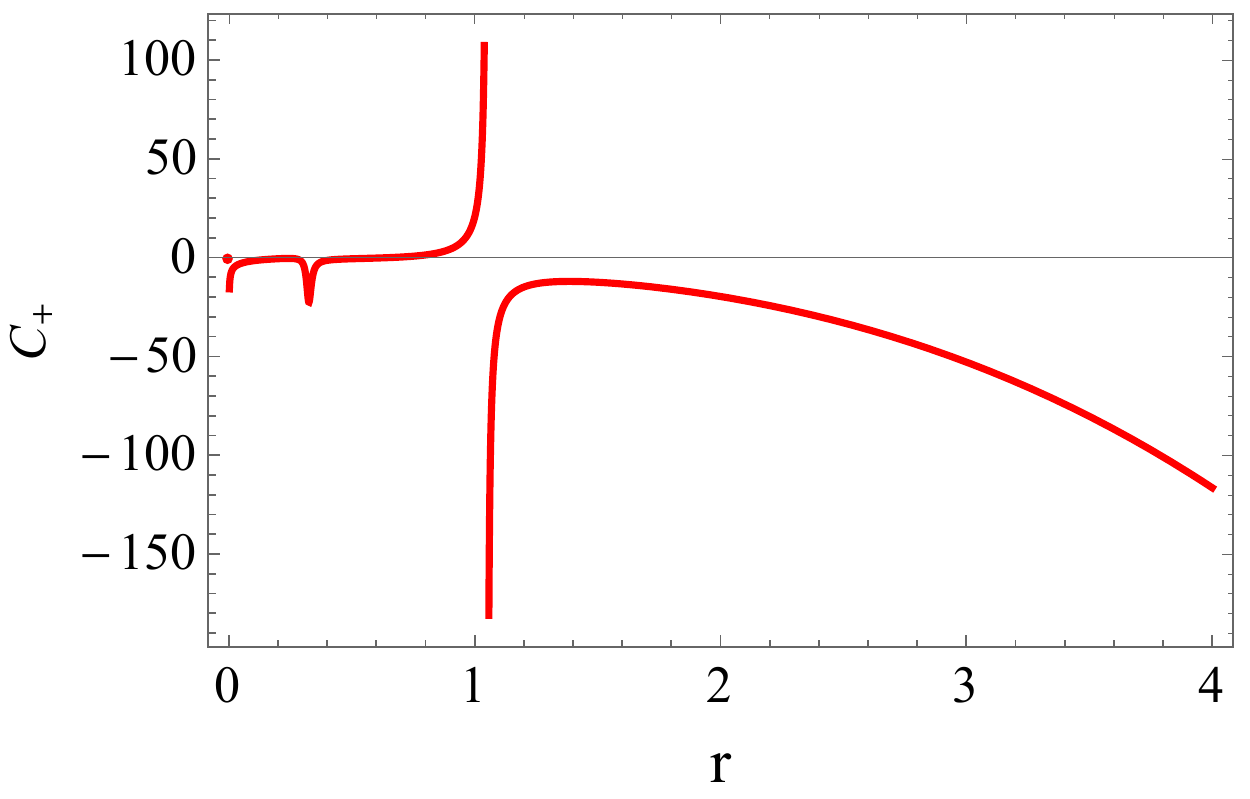}
\includegraphics[width=8.0 cm]{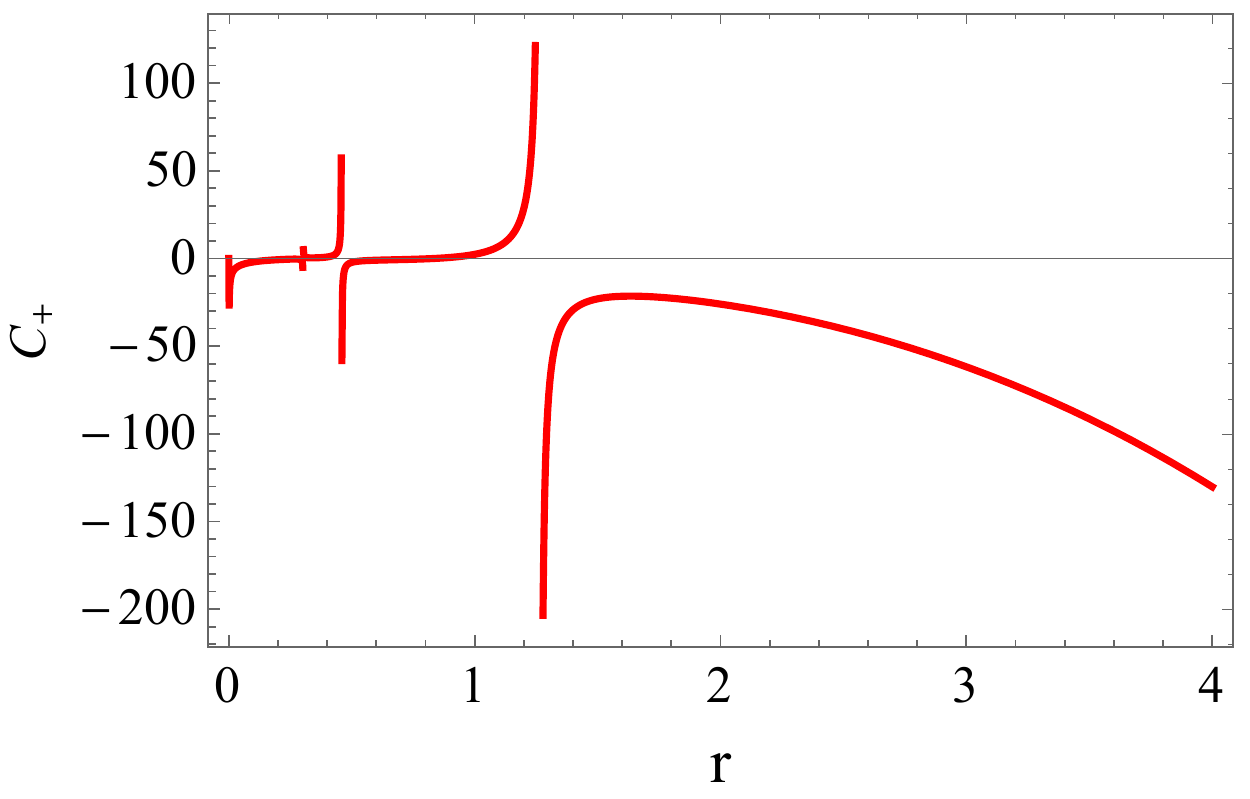}
\caption{Upper left panel: Plot for $C_{+}$ as a function of $r=r_+$ for chosen $\alpha=0.1$, $\beta=0.1$, $M=1$ and $q=0.2$. Upper right panel: Plot for $C_{+}$ for chosen $\alpha=0.1$, $\beta=0.1$, $M=1$ and $q=0.4$. Button left panel: Plot for $C_{+}$ for chosen $\alpha=0.1$, $\beta=0.1$, $M=1$ and $q=0.6$. Button right panel: Plot for $C_{+}$ for chosen $\alpha=0.1$, $\beta=0.1$, $M=1$ and $q=0.8$.   }
\end{figure*}
\begin{figure*}
\includegraphics[width=8.0 cm]{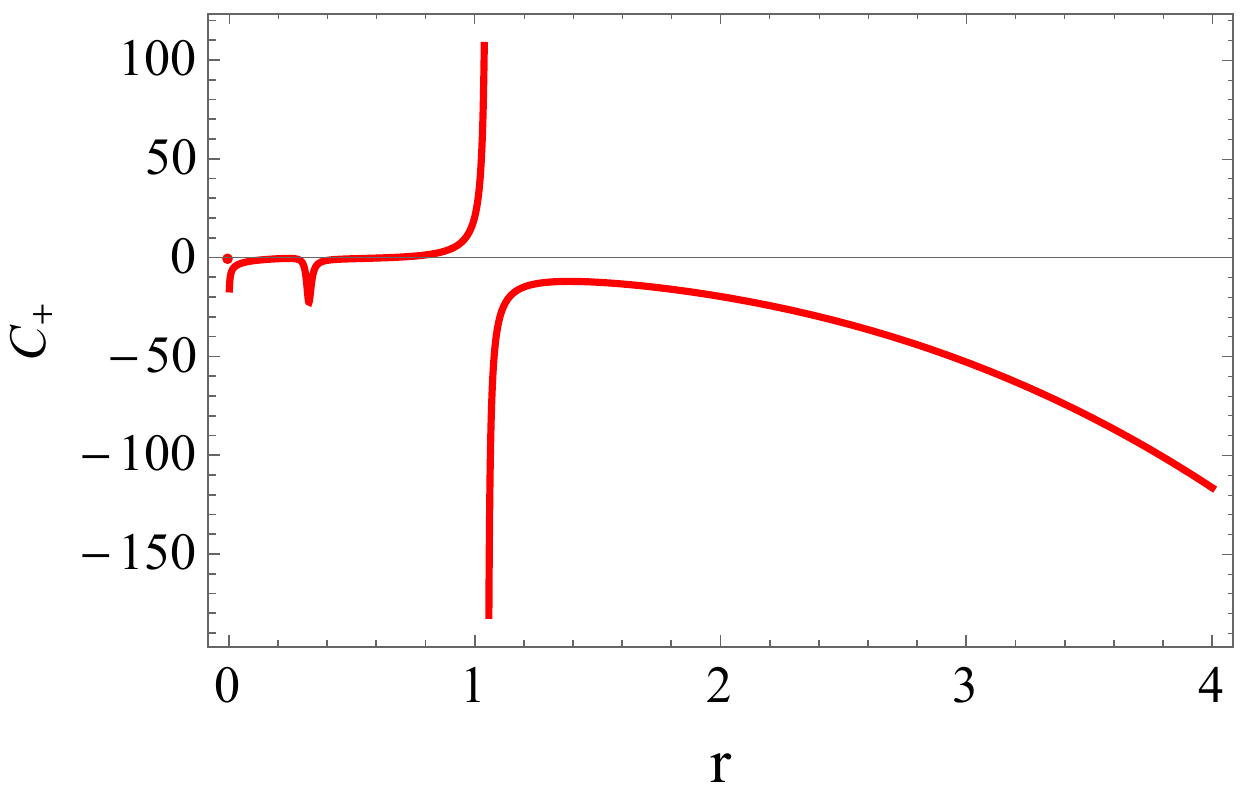}
\includegraphics[width=8.0 cm]{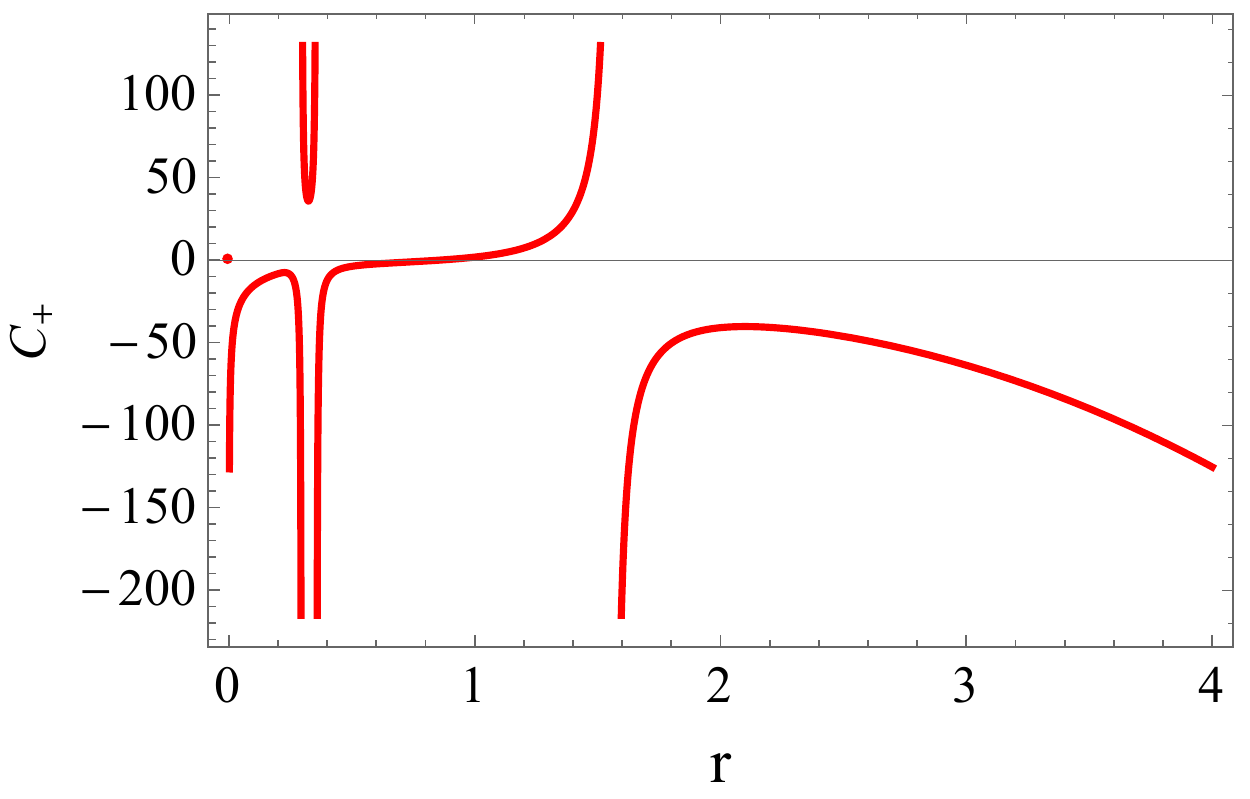}
\includegraphics[width=8.0 cm]{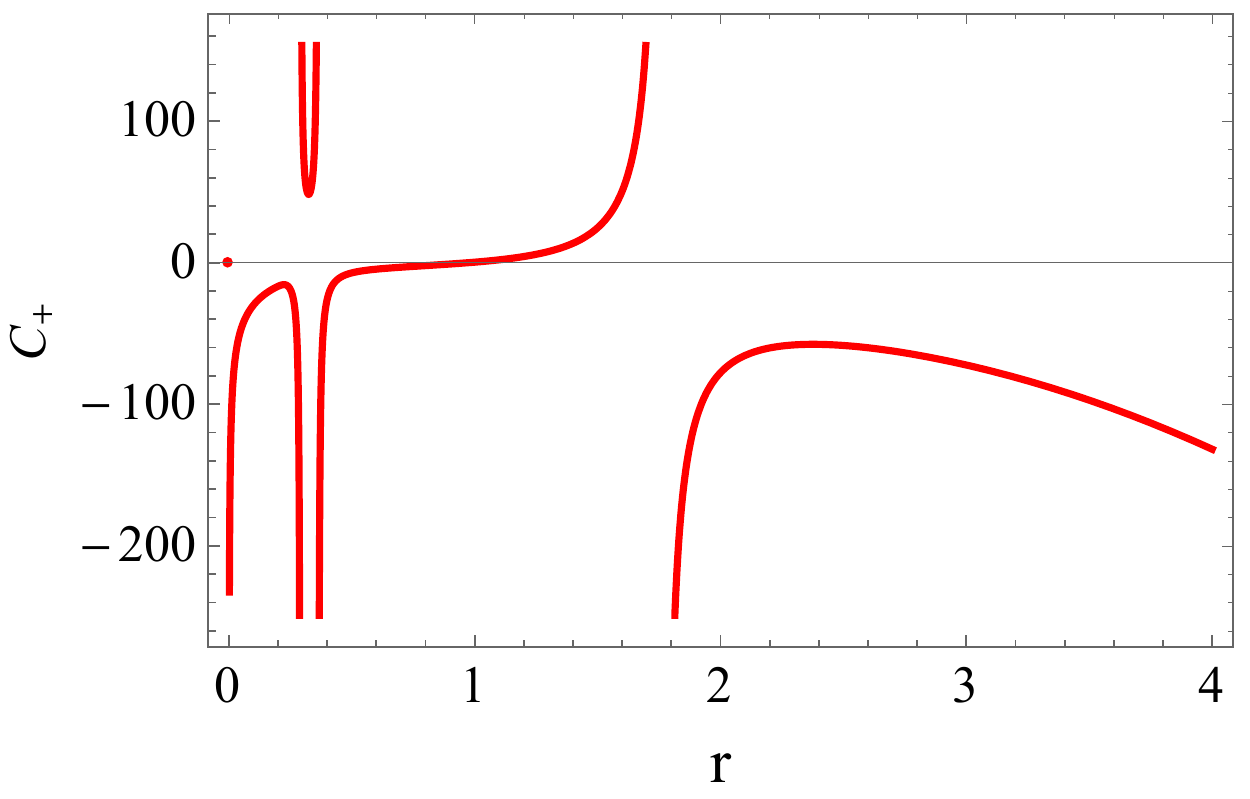}
\includegraphics[width=8.0 cm]{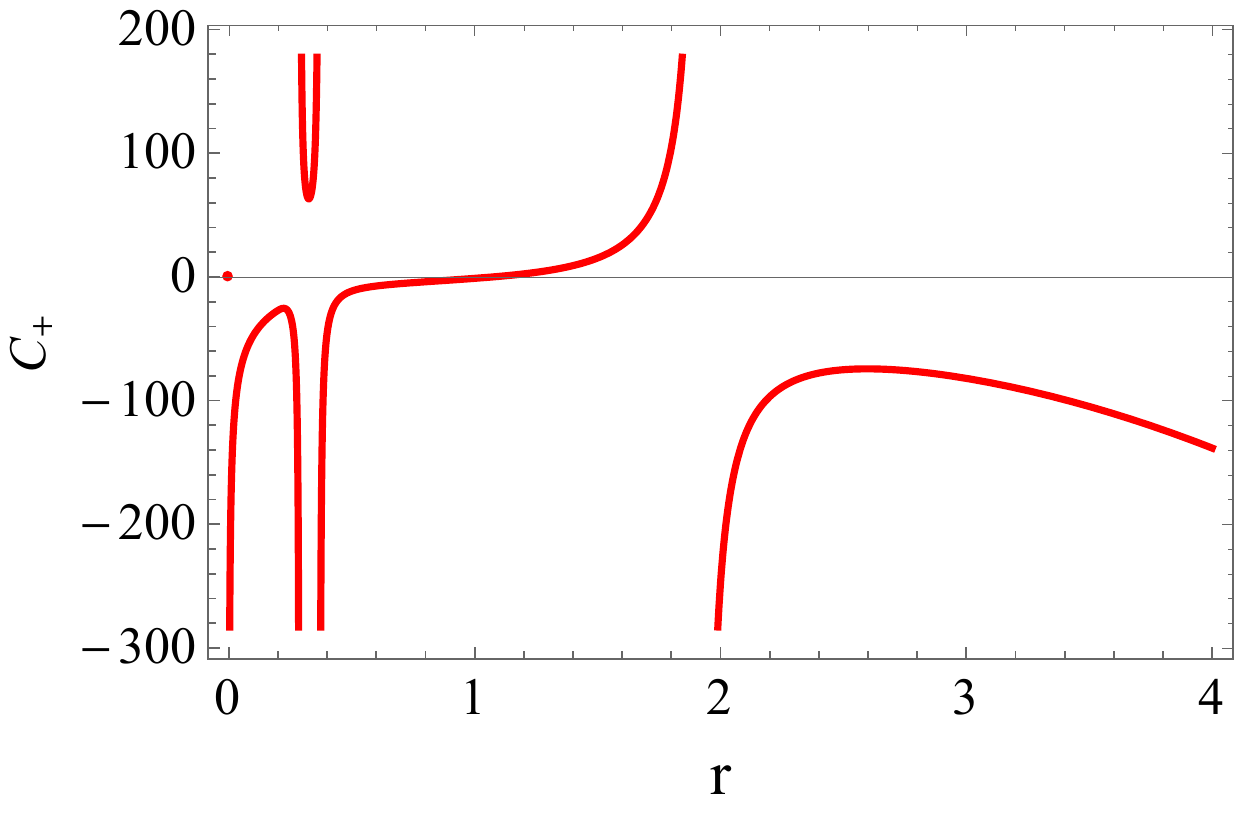}
\caption{Upper left panel: Plot for $C_{+}$ as a function of $r=r_+$ for chosen $\alpha=0.1$, $\beta=0.1$, $M=1$ and $q=0.6$. Upper right panel: Plot for $C_{+}$ for chosen $\alpha=0.4$, $\beta=0.1$, $M=1$ and $q=0.6$. Button left panel: Plot for $C_{+}$ for chosen $\alpha=0.6$, $\beta=0.1$, $M=1$ and $q=0.6$. Button right panel: Plot for $C_{+}$ for chosen $\alpha=0.8$, $\beta=0.1$, $M=1$ and $q=0.6$.   }
\end{figure*}

Taking the limit $\beta \to 0$ we obtain
\begin{equation}
\lim_{\beta \to 0}f(r)=1+\frac{r^2}{2 \alpha}\left(1 \pm \sqrt{1+\frac{8 M \alpha}{r^3}-\frac{4 q^2 \alpha}{r^4}}\right)
\end{equation}
which is the charged solution in 4D EGB with a vanishing cosmological constant reported in Ref. [5]. The $\pm$ sign in Eq. (14) refers to two different branches of
solution. Boulware and Deser \cite{Boulware:1985wk} have demonstrated that EGB black
holes with $+$ve branch sign are unstable and the graviton degree of freedom is a ghost, while the branch with $-$ve sign is stable and is free of ghosts. 
In our case, in the limit $\alpha \to 0$, the  $+$ve positive branch leads to 
\begin{widetext}
\begin{equation}
  f(r)=\frac{r^2}{\alpha}+\frac{2M}{r}-\frac {q^2 \exp\left(-\frac{\beta q^2}{2 r^4} \right)[4 \text{WhittakerM}\left(\frac{1}{8},\frac{5}{8},\frac{q^2 \beta}{2r^4})\right) \exp\left(\frac{\beta q^2}{4 r^4} \right)2^{1/8}+5(\frac{q^2 \beta}{r^4})^{1/8}]}{5 (\frac{q^2 \beta}{r^4})^{1/8}r^2}+ \ldots, \ldots
\end{equation}
which is a wormhole solution in a de-Sitter/ anti-de Sitter spacetimes depending on the sign of $\alpha$. On the other hand, in the limit $\alpha \to 0$, the  $-$ve  goes over
\begin{equation}\notag
f(r)=1-\frac{2M}{r}+\frac {q^2 \exp\left(-\frac{\beta q^2}{2 r^4} \right)[4 \text{WhittakerM}\left(\frac{1}{8},\frac{5}{8},\frac{q^2 \beta}{2r^4})\right) \exp\left(\frac{\beta q^2}{4 r^4} \right) 2^{1/8}+5(\frac{q^2 \beta}{r^4})^{1/8}]}{5 (\frac{q^2 \beta}{r^4})^{1/8}r^2}+ \ldots,
\end{equation}
 \end{widetext}
From the last two equation we can perform the limit $\beta \to 0$ to find find 
\begin{equation}
f(r)=\frac{r^2}{\alpha}+\frac{2M}{r}-\frac {q^2}{r^2}+ \ldots,
\end{equation}
for the $+$ve branch sign and 
\begin{equation}
f(r)=1-\frac{2M}{r}+\frac {q^2}{r^2}+ \ldots,
\end{equation}
the $-$ve branch sign, respectively. The last result is nothing but the charged black hole solution of GR. In that sense, if $q$ is replaced by the electric charge our solution (14) is a generalization of the recent work presented in Ref. \cite{Fernandes:2020rpa} when the cosmological constant vanishes. We notice that in Refs. \cite{Fernandes:2020nbq,Hennigar:2020lsl} a well well defined $D \to 4$ limit of EGB gravity was presented. Importantly in the case of
spherically symmetric spacetimes in $4D$ our black hole solution should remain valid in these regularised theories, however by going
beyond the sphericaly symmetric cases the solutions are not valid. 

\section{Embedding Diagram}
	In this section, we shall explore the geometry of our black hole solution by embedding it into a higher-dimensional Euclidean space. To simplify the problem let us consider the  equatorial plane $\theta=\pi/2$ at a fixed moment  $t=$ Constant, in that case we have
	\begin{equation}
	ds^2=\frac{dr^2}{1-\frac{b(r)}{r}}+r^2d\phi^2, \label{emb}
	\end{equation}
	where
	\begin{equation}
	b(r)= r(1-f(r)),
	\end{equation}
	Let us embed this black hole metric into three-dimensional Euclidean space  in the cylindrical coordinates,
		\begin{eqnarray}
		ds^2&=&dz^2+dr^2+r^2d\phi^2
		\end{eqnarray}
		From Eqs. (20) and (22),  we find that
	\begin{equation}
	\frac{dz}{dr}=\pm \sqrt{\frac{r}{r-b(r)}-1},
	\end{equation}
	where $b(r)$ is given by Eq.(21). Note that the integration of the last expression cannot be accomplished analytically. Invoking numerical techniques allows us to illustrate the embedding diagrams given in Fig. 2.

\section{Black hole thermodynamics}
In this section we shall discuss the thermodynamical properties of our magnetically charged black hole solution. Toward this goal we first compute the
gravitational mass of a black hole by solving $f(r_+)=0$, yielding
\begin{widetext}
\begin{equation}\notag
M(r_+)=\frac {2 \exp\left(-\frac{\beta q^2}{2 r^4} \right)[2^{1/8} q^2 \text{WhittakerM}\left(\frac{1}{8},\frac{5}{8},\frac{q^2 \beta}{2r^4})\right) \exp\left(\frac{\beta q^2}{4 r^4} \right) 2^{1/8}+\frac{5((r^2+\alpha)\exp\left(\frac{\beta q^2}{2 r^4} \right)+q^2)}{4}(\frac{q^2 \beta}{r^4})^{1/8}]}{5 (\frac{q^2 \beta}{r^4})^{1/8}r^2}|_{r_+}
\end{equation}
\end{widetext}

In particular if we now take the limit $\beta \to 0$, we obtain
\begin{equation}
\lim_{\beta \to 0}M(r_+)=\frac{q^2+r^2+\alpha}{2 r}|_{r_+}
\end{equation}
a well known result. The Hawking temperature associated to our black hole solution can be found by using the relation
\begin{equation}
T_H=\frac{f'(r)}{4 \pi}|_{r_+}.
\end{equation}

Due to the complicated and long expression for $T_{H}$, in Fig. 3 we show the plots of Hawking temperature as a function of $r=r_+$. The existence of negative temperatures is unphysical from the thermodynamics point of view. Therefore, having negative temperature, simply means that the black hole is absent. However note that negative specific heats are a familiar feature of self-gravitating systems. In that sense, the heat capacity of the black hole can be negative. In particular, it is well known that due to thermal fluctuations, at some stage, a black hole can absorb more radiation than it emits which
leads to positive heat capacity. Whereas when the black hole emits more radiation than it absorbs, the heat capacity
becomes negative. Put in other words, we can say that the black hole is thermodynamically stable (unstable) to the thermal
fluctuations, if the heat capacity is positive (negative).  In our black hole solution one can show that there exists always a domain of parameters such that the Hawking radiation is negative. We have estimated some numerical values of parameters related to this issue.  As a particular example we consider a black hole with mass $M=1$, a horizon radius $r_+=1$, along with with a fixed nonlinear parameter $\beta=0.1$. For such a domain we find that by increasing the magnetic charge: $q=\{0.1, 0.2, 0.3, 0.4, 0.5\}$, we find that the black hole is absent in the following intervals: $\alpha_c= \{0.990007, 0.960109, 0.910537,0.841623, 0.753734\}$, respectively. Using the above parameters in the interval $\alpha>\alpha_c$, one can check that the function $f(r)$ evaluated at $r_+$ is positive i.e., $f(r_+)>0$, meaning that the there are no horizons at all. This confirms the fact that the black hole is absent in the interval $\alpha>\alpha_c$.  In other words, the smaller the value of the magnetic charge the greater the value of the Gauss-Bonnet parameter, hence the interval is larger, provided  $\alpha>0$.
As we already pointed out, the thermodynamical stability of the black hole can be found by using the heat capacity $C_+$. The stability of the black hole is related to sign of the heat capacity. In particular when $C_+>0$, the black hole
is stable, while in the case $C_+ <0$, the black hole is unstable. The heat capacity of the black hole is given by \cite{Singh:2020nwo}
\begin{equation}
C_+=\frac{\partial M_+}{\partial T_+}=\frac{\partial M_+}{\partial r_+}\frac{\partial r_+}{\partial T_+}.
\end{equation}

Again due to the long and complicated expression, in Figs. 4 and 5 we plot the heat capacity of the black hole as a function of $r=r_+$ for different values of parameters. In Fig. 4 we keep $\alpha$ constant and we increase the magnetic charge $\beta$. In Fig. 5 on the other hand we keep constant the magnetic charge $q$ and increase $\alpha$, respectively. It is observed that in general the heat capacity $C_+$ exhibits discontinuous and diverges at two critical points  $r=r_c$, which can be linked to the second-order phase transition. In particular there is a flip of sign in the heat capacity around $r_c$ where the Hawking temperature attains a maximum value with $(\partial T_+/\partial r_+)=0$. The black hole is thermodynamically stable for $r_+<r_c$ whereas it is thermodynamically unstable for $r_+>r_c$. We also see that in general, for large horizon radius $r_+$, we have negative heat capacity $C_+<0$ hence thermodynamically unstable. However, it is interesting to see that at late stages of the black hole evaporation the heat capacity is also negative i.e., $C_+<0$. Hence smaller mass black holes are also thermodynamically unstable. The black holes appears to the thermodynamically stable in some intermediate mass region with $r_+$ between smaller unstable region and larger unstable region. In fact, we can see from Figs. 4 and 5 that there exists a domain of parameters such that the black hole undergoes a phase transition twice. Firstly, the phase transition occurs at some $r_{c}^a$ from a higher mass unstable black hole with negative heat capacity $C_+<0$, to some intermediate stable region of the black hole mass with positive heat capacity $C_+ >0$. Secondly, there is a phase transitions at some $r_c^b$ from the stable intermediate black hole mass region with positive heat capacity to smaller unstable black hole with negative heat capacity $C_+<0$.

\section{Conclusion}
In this work we have found an exact solution of magnetically charged black holes with exponential model of nonlinear electrodynamics in the context of  4D EGB gravity.  
We have shown that our $-$ve branch results, in the limit $\alpha \rightarrow 0$ and $\beta \to 0$, reduced exactly  to the well known magnetically charged black hole of GR. We have analyzed the black hole geometry by embedding into three-dimensional Euclidean space.  We have also explored the thermodynamic properties such as the Hawking temperature and thermal stability of regular black holes.  It found that that the heat capacity diverges in two points $r=r_c$ hence the black hole undergoes a phase transition more then once. We have shown that for large horizon radius $r_+>r_c$, we have negative heat capacity $C_+<0$, meaning higher mass black holes are  thermodynamically unstable. Similarly, at late stages, due to the black hole evaporation we have found that the heat capacity is negative, suggesting that smaller mass black holes are also thermodynamically unstable. This shows that the black hole appears to the thermodynamically stable in the intermediate mass region having $r_+^a<r_+<r_+^b$, located between smaller unstable region and larger unstable region. We notice that in Ref. [36] a well well defined $D \to 4$ limit of EGB gravity and 
the spherically symmetric $4D$ black hole solution should remain valid in these regularised theories, but not
beyond spherical symmetry.

\end{document}